# Imaging three-dimensional nanoscale magnetization dynamics

**Authors:** Claire Donnelly[1,2,3*], Simone Finizio[2], Sebastian Gliga[2,3,4], Mirko Holler[2], Aleš Hrabec[2,3,5], Michal Odstrčil[2], Sina Mayr[2,3], Valerio Scagnoli[2,3], Laura J. Heyderman[2,3], Manuel Guizar-Sicairos[2*], Jörg Raabe[2*].

**Affiliations:**

[1] Cavendish Laboratory, University of Cambridge, JJ Thomson Ave, Cambridge, CB3 0HE, UK.

[2] Paul Scherrer Institute, 5232 Villigen, Switzerland.

[3] Laboratory for Mesoscopic Systems, Department of Materials, ETH Zurich, 8093 Zurich, Switzerland.

[4] SUPA, School of Physics and Astronomy, University of Glasgow, Glasgow G12 8QQ, UK.

[5] Laboratory for Magnetism and Interface Physics, Department of Materials, ETH Zurich, 8093 Zurich, Switzerland

*Correspondence to: cd691@cam.ac.uk, manuel.guizar-sicairos@psi.ch, joerg.raabe@psi.ch.

**Abstract:** The ability to experimentally map the three-dimensional structure and dynamics in bulk and patterned three-dimensional ferromagnets is essential both for understanding fundamental micromagnetic processes, as well as for investigating technologically-relevant micromagnets whose functions are connected to the presence and dynamics of fundamental micromagnetic structures, such as domain walls and vortices. Here, we demonstrate time-resolved magnetic laminography, a technique which offers access to the temporal evolution of a complex three-dimensional magnetic structure with nanoscale resolution. We image the dynamics of the complex three-dimensional magnetization state in a two-phase bulk magnet with a lateral spatial resolution of 50 nm, mapping the transition between domain wall precession and the dynamics of a uniform magnetic domain that is attributed to variations in the magnetization state across the phase boundary. The capability to probe three-dimensional magnetic structures with temporal resolution paves the way for the experimental investigation of novel functionalities arising from dynamic phenomena in bulk and three-dimensional patterned nanomagnets.

**One Sentence Summary:** Nanoscale dynamics of a three-dimensional magnetic system are revealed with pump-probe magnetic laminography.

**Main Text:**

Understanding and controlling the dynamic response of magnetic materials with a three-dimensional magnetization distribution is important both fundamentally and for technological applications. From a fundamental point of view, the internal structure and reversal phenomena in known bulk materials still need to be mapped [1], along with the dynamic properties of topological structures such as vortices [2], magnetic singularities [3] or skyrmion lattices [4]. From a technological point of view, the response of inductive materials to magnetic fields and spin-polarized currents is essential for magnetic sensors and data storage devices [5].

In addition to bulk systems, curved and three-dimensional magnetic structures such as nanotubes and nanowire-based systems have been predicted to exhibit dynamic properties markedly different from those found in planar systems. These include the suppression of the Walker breakdown in magnetic nanotubes and cylindrical nanowires and fast domain wall velocities measured in three-dimensional magnetic microwires [6], as well as anisotropic spin wave propagation [7] and the spin-Cherenkov effect [8, 9].

While magnetization dynamics have been observed in 2D structures, and are well understood for isolated structures in thin films such as vortices [10-13] domain walls [14-17] and skyrmions [18, 19], until now insight into three-dimensional dynamics has only been possible via comparison with simulations [20].

Static three-dimensional magnetic imaging has recently been achieved using hard X-ray [21, 22], neutron [23-25], electron [26] and soft X-ray tomographies [27, 28] which allow for the determination of the three-dimensional magnetic configurations. Of these methods, X-rays are unique owing to their ability to provide high spatial resolution imaging of thick materials [21, 29], as well as element-specific information. Additionally, the pulsed time structure of synchrotron X-rays provides the possibility of performing time-resolved measurements: pump-probe measurements of material systems have proved a powerful tool for the measurement of MHz and GHz dynamics within a range of magnetic materials and systems [30]. As yet, however, it has not been possible to combine three-dimensional magnetic imaging with dynamic measurements due to the geometric constraints of tomographic imaging on sample design.

Here we present an alternative geometry for 3D magnetic imaging: magnetic laminography [31], that provides access to all three components of the magnetization vector field with a single axis of rotation, and is a flexible technique that makes possible the determination of the temporal evolution of three-dimensional magnetic configurations. By combining vectorial magnetic imaging with a pump-probe setup, and therefore directly observing magnetization dynamics of complex three-dimensional magnetic configurations, we obtain access to the temporal evolution of the entire three-dimensional vector field. In this way, we can study both surface excitations as well as the dynamics of buried three-dimensional magnetization textures such as magnetic domains and vortex domain walls.

We specifically image the three-dimensional magnetic state of a GdCo microdisc using magnetic laminography, and determine its dynamic response to a radio frequency (RF) magnetic field. The sample consists of a GdCo microdisc with a diameter of 5 μm and a thickness of 1.2 μm, patterned by focused ion beam milling a continuous film deposited on the back-side of a $Si_3N_4$ membrane (see inset of Fig. 1a for an SEM image of the sample). The two-phase GdCo film was deposited by magnetron sputtering, first with the sample continuously rotating, and then stationary for the top 1/4 of the thickness. On the front-side of the membrane a 300 nm-thick Cu stripline was patterned, through which an RF current could be injected to produce an in-plane magnetic excitation.

The experimental setup is shown schematically in Figure 1a, where the plane of the sample is tilted at 29° to the propagation direction of the X-rays. During the measurement the sample is rotated about an axis perpendicular to its surface plane, and two-dimensional projections of the magnetic structure are measured with dichroic ptychography [29]. Two-dimensional projections of the magnetic configuration of the GdCo microdisc for the sample rotated by 0° and 45° about the laminography rotation axis are shown in Figure 1b and 1c, respectively. In these projections, a multi-domain state can be identified at the bottom of the structure, that appears to change through the thickness of the structure.

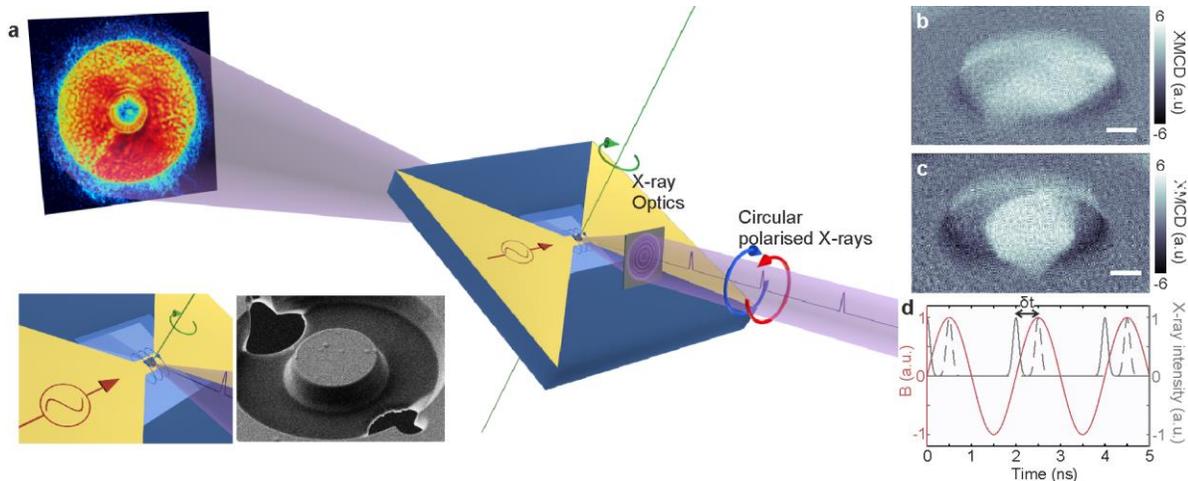

Figure 1. Time resolved X-ray magnetic laminography. a) the experimental setup. For X-ray magnetic laminography, ptychographic projections are measured with right- and left-handed circular polarization with the sample rotated around an axis, that is at 61° with respect to the X-ray beam propagation direction. To probe the magnetization dynamics of the sample, the structure is fabricated on a Cu stripline (see magnified view in inset, along with an SEM image of the structure), and excited with a 500 MHz RF current passing through the stripline. Magnetic laminography is then performed for a number of different delay times of the excitation with respect to the X-ray pulses, so providing temporal resolution. XMCD projections of the magnetic structure are shown in b) and c) for the sample at rotation angles of 0° and 45°, respectively. Scale bars represent 1 μm. d) the concept of pump-probe magnetic laminography is illustrated, where the RF excitation is frequency and phase matched to the incoming X-ray pulses. By varying the time delay δt for a number of laminography measurements (two are depicted here), temporally-resolved maps of the magnetization vector field are obtained.

For magnetic laminography, 2D ptychographic projections were measured at 144 angles equally distributed over 360° with both right and left circularly polarized X-rays. The three-dimensional magnetic configuration was reconstructed using a GPU implementation of the gradient-based arbitrary projection magnetic reconstruction algorithm (APMRA), detailed in [32], with a lateral spatial resolution of ~50 nm. The RF magnetic field used to excite the sample was frequency and phase-matched to the synchrotron X-ray pulses, which are 70 ps wide X-ray flashes occurring every ~2 ns (repetition frequency of 500 MHz). To probe the dynamic response of the sample, a magnetic laminography dataset was acquired for a number of different delay times of the excitation field with respect to the X-ray bunches, as shown schematically in Figure 1d.

The reconstructed three-dimensional magnetization distribution of the GdCo microdisc is shown in Figure 2. In the lower region of the structure, a multi-domain state is present, in which the domains are separated by elongated vortex domain walls (Figure 2a). As the disk thickness increases, the central domain expands and the vortex domain walls are located increasingly closer to the edges of the sample (Figure 2b), until they are expelled approximately mid-way through the thickness of the sample, resulting in a single magnetic domain with a curling magnetization at the edges, forming an S-state (Figure 2c). At the top of the disc, the magnetization forms an almost uniform state (Figure 2d). The positions of the vortex cores are plotted using isosurfaces in Figure 2e. Within the core of a vortex, the magnetization points out of the plane of the disk, and the direction of the magnetization (i.e. $\pm m_y$), defines its *polarization*. Here, the isosurfaces representing the core of the vortices are colored by the $m_y$ component, which can be seen to be positive (orange) and negative (green), indicating that the pair of vortices have opposite polarizations.

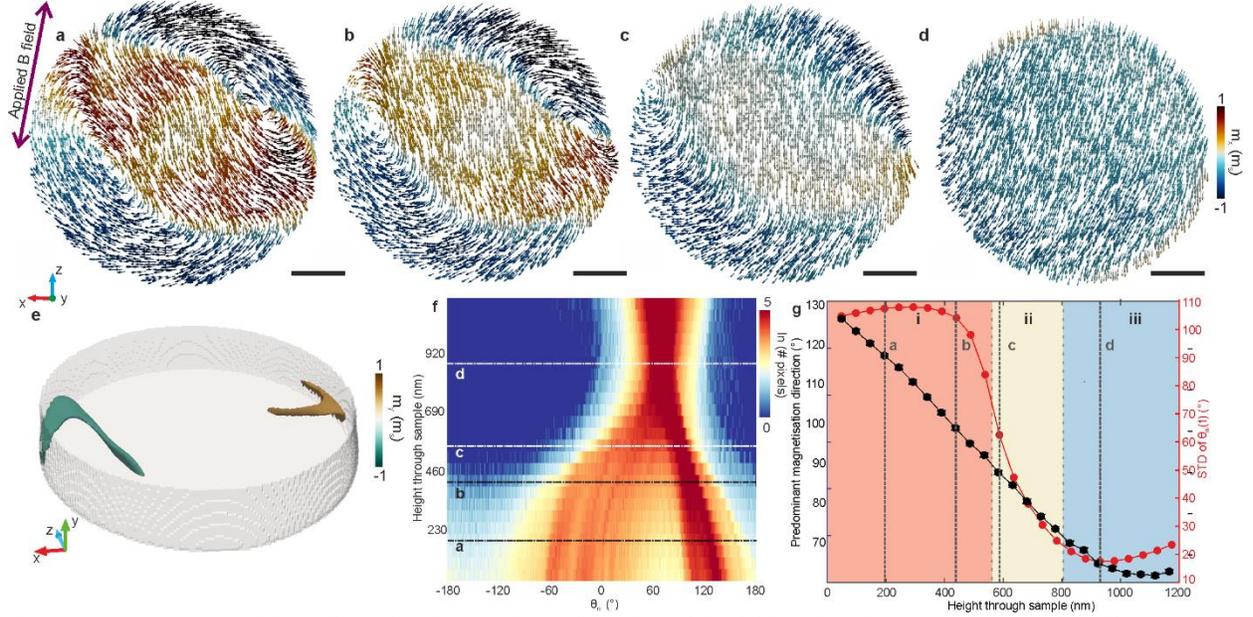

Figure 2 The reconstructed three-dimensional magnetic configuration of the GdCo microdisc. At the bottom of the disc (a), an "eye" state forms with a large central domain, and two vortices. With increasing height, the vortices move towards the edges of the disc (b), until they exit the structure approximately midway through the thickness, resulting in a large magnetic domain in an S-state (c). At the top of the disc, the magnetic configuration is almost uniform (d). The positions of the cores of the vortices, coloured by the y-component of the magnetization, are shown in (e), showing that the magnetization in the cores of the two vortices have opposite directions – i.e. opposite vortex polarizations. These vortices eventually exit through the edges of the sample approximately mid-way through its height. f) Bivariate histogram showing the number of pixels as a function of magnetization direction and thickness through the sample. The red peak corresponds to the central domain, which expands in area with increasing height, eventually forming a single-domain state above ~700 nm. Three main regions in the histogram are identified: in the vortex state (h≤500 nm, a,b), there is a large variation in magnetization direction, with two main peaks corresponding to the central and side domains. At h ~500 nm, an intermediate S-domain state is present (c), and the magnetization directions are distributed at angles between ~50 and 90°. Above 800 nm, the single domain state (d) results in a strong, sharper peak around 65°. g) The primary direction of the magnetization in the central domain (black dots) and the standard deviation of $\Theta_m$ (red dots), which indicates the magnetization state, are given as a function of thickness through the sample. The primary direction varies linearly up to a thickness of ~900 nm, after which the direction remains approximately constant, consistent with the presence of in-plane magnetic anisotropy in the top quarter of the film. Three regions of the sample are identified on the plot, the vortex state (i, orange), the S-state (ii, yellow) and the single domain state (iii, blue).

The gradual transition between the multi-domain state in the lower region of the sample to the single domain state in the upper region can be attributed to variations in the anisotropy of the system. The rotation of the sample during sputtering prevented the formation of a well-defined anisotropy axis in the bottom part of the disk so that the magnetic state is predominantly determined by the magnetostatic energy. The absence of rotation during the latter part of the deposition led to the development of a sizeable uniaxial anisotropy in the top part of the sample. Therefore, the resulting system can be understood as a two phase system with a soft (bottom) and a hard (top) phase, conceptually analogous to an exchange-spring system [33].

We compare the distribution of the direction of the magnetization through the thickness of the sample in Figure 2f in a bivariate histogram. At the bottom of the film, there is a large spread in the magnetization directions, which is consistent with the vortex state, with a distinct peak at 140° corresponding to the direction of the magnetization in the central domain. With increasing height, the spread in directions is reduced, giving rise to a single peak corresponding to the direction of the magnetization in the top domain. The standard deviation of the magnetization, which reflects the magnetization state, is plotted as a function of thickness through the sample in Figure 2g (red dots), and three main regions can be identified: the vortex state (i), the single domain state (iii), and an intermediate region corresponding to the curled S-state (ii). The primary direction of the magnetization can be tracked through the thickness of the sample by identifying the corresponding peak in the histogram for each slice in Figure 2g (black dots). The primary magnetization direction varies linearly with distance through the sample up to a height of 900 nm, after which the magnetization direction remains almost constant. This linear rotation of the magnetization direction along the thickness correlates with the position of the vortex walls: as the thickness increases, the vortices (plotted using isosurfaces in Fig 2e) are located increasingly closer to the edges of the sample and eventually expelled with the onset of strong uniaxial anisotropy in the top part of the disk.

We now probe the magnetization dynamics in response to the RF excitation field of amplitude ~4 mT, applied in the direction shown in Figure 2, by performing magnetic laminography for seven delay times over a period of 2 ns with respect to the X-ray pulse train.

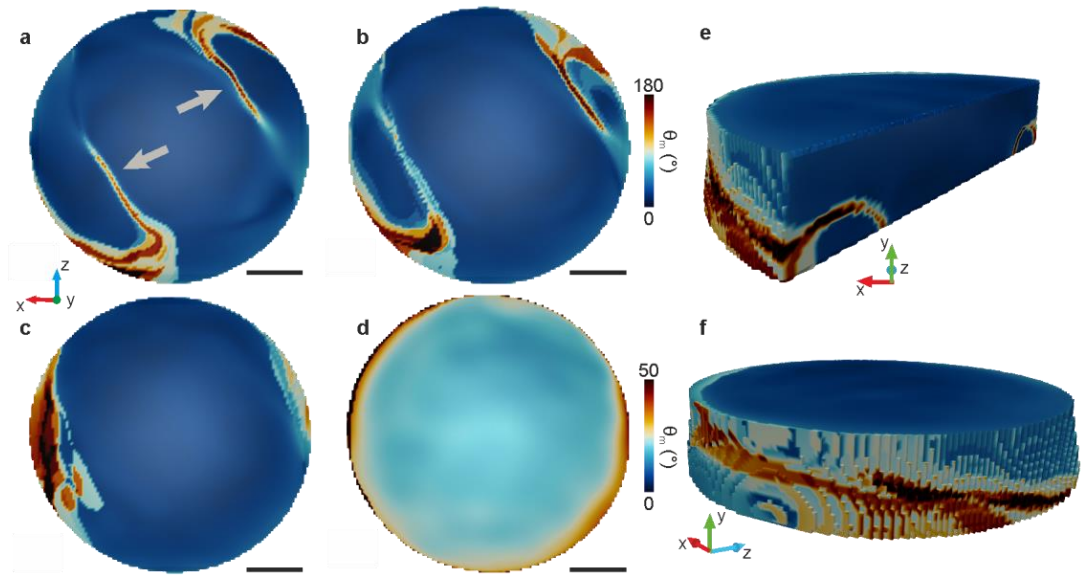

Figure 3. Dynamic response of the three-dimensional magnetization at 500 MHz. a) The angle at which the magnetization precesses is plotted for the heights within the sample corresponding to Fig 2 a-d (see Fig. S4 for slice locations). In the vortex state (a,b) there is significant precession in the vicinity of the vortex domain walls (indicated with white arrows in a), as well as at the edge of the sample, indicating the presence of pronounced edge modes. In the intermediate S-state (c), an enhancement of the edge modes at opposite sides of the disc is observed. In the single domain state (d), the precession is significantly reduced, however lower amplitude edge modes are present on the top left and bottom right edges of the structure (this is visualised with a different colour scale to a-c,e,f.) The three-dimensional dynamics are visualised in e, and f, by mapping the magnetisation precession, where one can see that the dynamics are limited to the vortex cores (e) and edge modes that spiral around the disc (f). Scale bars represent 1 μm.

We map the spatial distribution of the magnetization dynamics by the Fourier coefficient of the temporal evolution of the in-plane magnetization direction $\theta_m$ corresponding to 500 MHz, in Figure 3. The precession angle of the magnetization is plotted for the four magnetic states shown in Figure 2a-d in Figure 3a-d, respectively. While the direction of the applied RF field is almost parallel to the direction of the magnetization in the top region of the disk, magnetization dynamics are observed across the entire thickness, as a result of the inhomogeneous magnetization structure. At the bottom of the sample (Fig. 3 a,b – region (i)) the magnetization oscillates in the vicinity of the vortex walls indicating their reversible displacement (red lines, indicated by white arrows in Fig. 3a). The magnetization also oscillates at the bottom left and top right edges of the disc, where the magnetization is almost perpendicular to the applied RF field, thereby maximizing the field-induced torque. In the intermediate S-state (region (ii)), magnetization precession is observed in two regions close to the sample edge, which are adjacent to the location where the vortices are expelled from the disc (Figure 3c). In the single domain state (Figure 3d) oscillations of the magnetization in the bulk of the domain are absent, and only lower amplitude edge modes are present (made visible by the more limited color scale). Three-dimensional maps of the modes are given in Figure 3e,f, showing that the magnetization dynamics are limited to the vortex domain walls and at the edges of the sample. Mapping the magnetization dynamics gives a picture of the interaction between the different domain structures and of the resulting modifications in their dynamics. Across the sample height, the edge oscillations are located at different positions, describing a spiral, which follows the evolution of the location of the vortex walls, due to the presence of the vortices giving rise to regions with a sizeable magnetization component perpendicular to the applied field direction, where the applied torque is greater. In addition, the large anisotropy in the top region would be expected to lead to the formation of an 'onion' state [34] characterized by dynamic edge modes at the extremities of the domain, where the magnetization typically is not parallel to the sample boundaries. In contrast, in our system, imprinting of the bottom vortices onto the top region rather leads to the formation of a 'flower' state [35], thus shifting the regions in which the magnetization oscillates.

Having identified the regions excited by the applied field, we now consider the details of the temporal evolution of the vortex domain walls in the lower section of the disc in Figure 4. The curl of the in-plane magnetization $\nabla \times m_{xz}$ is plotted for the slice at h=90 nm in Figure 4a, where the red and blue stripes correspond to the two vortex walls, which have opposite chiralities. In the insets, $\nabla \times m_{xz}$ is plotted in the vicinity of the domain wall with positive vorticity (region indicated by the black dashed box) for different time delays spanning a period of 2 ns. Over this period, the domain wall moves about the reference position (dashed line) and its displacement in the direction perpendicular to the long axis of the wall is plotted as a function of time for $h = 46$ nm in Figure 4b, and for all sample heights in Supplementary Figure S6, where the vortex domain wall can be seen to consistently exhibit an oscillatory motion. The amplitude of the domain wall oscillations is determined by calculating the Fourier coefficient of the temporal domain wall displacement to be 74±4 nm over the bottom 500 nm of the sample, with a maximum domain wall displacement of 200±15 nm throughout the structure, corresponding to a domain wall velocity of approximately 200 m/s.

The domain wall with negative vorticity (blue in Figure 4a) displays a similar oscillatory motion perpendicular to its long axis with an amplitude of (45±13) nm. Plotting the displacement of the domain walls as a function of time (Figure 4b) reveals that the domain walls oscillate out of phase. We quantify the phase difference between the two domain walls by calculating the phase of the Fourier coefficients, and find it to be $\frac{3\pi}{4} \pm \frac{\pi}{4}$, equivalent to a delay of $0.75 \pm 0.25$ ns. This out-of-phase motion corresponds to the breathing of the central domain, which is due to the torques exerted by the field on the two walls with opposite vorticities [12].

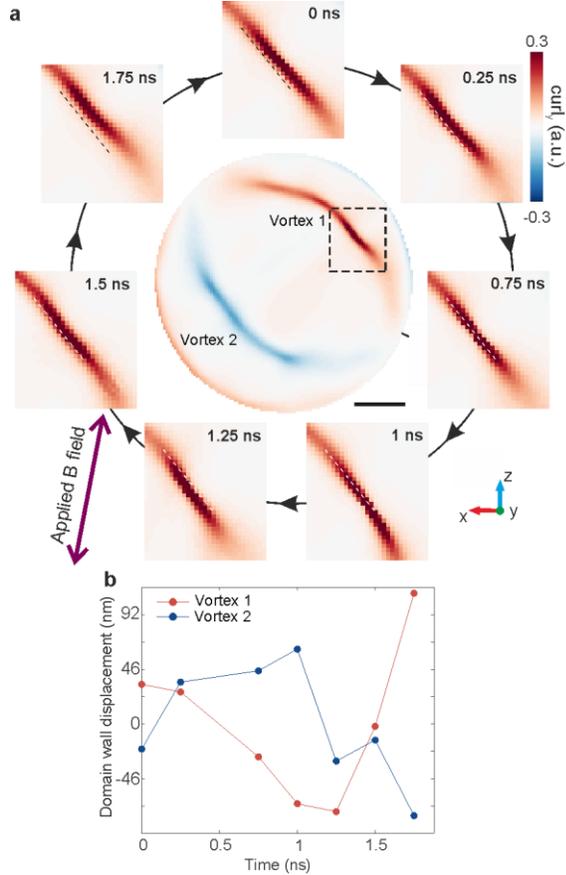

Figure 4. Domain wall dynamics in the lower region of the disc containing two vortex walls. a) The curl of the in-plane magnetization $\nabla \times m_{xz}$ for h = 90 nm is shown, where the two domain walls are identified by the positive (red) and negative (blue) lines. The domain wall with positive curl in the region indicated by the dashed box is shown at different times increasing clockwise around the central image, with a reference location indicated by a dashed line. b) The dynamics of the wall with a positive curl vortex (red, vortex 1) are compared with the dynamics of the wall with a negative curl vortex (blue, vortex 2) for an example slice of height 46 $nm$. The two oscillatory signals are out of phase by $(0.75 \pm 0.25)$ $ns$, consistent with the "breathing" of the central domain with the applied RF field. Scale bar represents 1 μm.

By combining magnetic laminography with a pump-probe experimental setup, we have probed the magnetization dynamics of a three-dimensional magnetic structure in response to the application of an RF magnetic field. Two main types of magnetization dynamics are identified: domain wall precession and magnetization dynamics at the edges of a uniform magnetic domain that are enhanced due to variations in the magnetization state across the phase boundary. At the boundary of two phases defined by different magnetic anisotropies, the local transfer of anisotropy from the top layer to the soft layer through exchange coupling leads to the expulsion of the domain walls and the enhancement of edge dynamics. Since laminography is ideal for the study of flat and thin samples, pump-probe laminography is directly compatible with high spatial resolution soft X-ray imaging, making possible the direct application of the technique to nanoscale systems. The capability to experimentally measure complex magnetization dynamics in three-dimensional systems is key for the investigation of promising dynamic properties predicted in such systems, that have been proposed for future data storage[36], logic [37] and microwave [38, 39] applications.

**Acknowledgments:**

The authors would like to thank Dario Marty for initial discussions about the sample fabrication and design, Elisabeth Müller for FIB patterning of the sample. All data was measured at the cSAXS beamline, Swiss Light Source, Paul Scherrer Institute, Switzerland.

**Funding:** C.D. acknowledges funding from the Leverhulme Trust (ECF-2018-016), the Isaac Newton Trust (18-08) and the L'Oréal-UNESCO UK and Ireland Fellowship For Women In Science 2019. A.H. was funded by the European Union's Horizon 2020 research and innovation programme under Marie Sklodowska-Curie grant agreement number 794207 (ASIQS).

**Author contributions:** C.D. and S.F. conceived the study of 3D magnetic dynamics with input from J.R.. The sample fabrication was planned by C.D. and S.F. and performed by C.D., A.H., S.M. and E.M. The time-resolved setup was created by S.F. and J.R. The laminography endstation was designed and built by M.H. The synchrotron measurements were performed by C.D., S.F., M.G.-S., M.H., V.S., and J.R.. The non-magnetic laminography reconstruction code was developed by M.O., and used by C.D. with the support of M.O. The magnetic laminography reconstruction was developed and used by CD, with the support of M.G-S. Data analysis was performed by C.D. with the support of M.G-S. The time-resolved 3D magnetic data was interpreted by C.D., S.F., S.G. and M.G.-S. CD wrote the manuscript with contributions from all authors.

**Competing interests:** Authors declare no competing interests.

**Data and materials availability:** All data and analysis code associated with this manuscript will be made available on a repository once the manuscript is published.

**Supplementary Materials:**

**Experimental setup**

*Laminography setup*

A gold Fresnel zone plate (FZP) with a diameter of 170 µm and an outermost zone width of 60 nm was coherently illuminated at a photon energy of 7.247 keV, which corresponds to the maximum XMCD signal at the Gd $L_3$ edge. The FZP was fabricated with displaced zones, designed to introduce perturbations into the illumination wavefront and improve the image resolution for ptychography [40]. The FZP was used in combination with a 50 $\mu$m-diameter central stop and 30 $\mu$m-diameter order sorting aperture. Circular polarized light was generated by a diamond quarter wave plate phase plate positioned before the setup as presented in [21, 29], with a resulting degree of polarization of over 99.9%, and an efficiency of approximately 60%.

The sample was placed downstream of the focal spot of the FZP where the beam size reached 5 $\mu$m in diameter. The sample was mounted on a 2D piezo scanner with 100 x 100 $\mu m^2$ scan range and the scanner was installed on a rotation stage with a clear aperture allowing 360° rotation. Two linear stages with a travel range of several millimeters were used to move the rotation stage in order to keep the region of interest in the field of view at different rotation angles. The entire sample stage stack was mounted at the laminography angle of 61° (angle between axis of rotation and X-ray beam propagation). The angle was chosen to provide a balance between 3D spatial resolution and sample absorption, and was checked with numerical simulations of magnetic laminography that are described in the next section.

Similar to previous instrumentation [41, 42], closed loop sample positioning and thus positioning accuracy during ptychographic scans was achieved by differential laser interferometry measuring the relative position between the sample and the beam-defining FZP in both the horizontal and vertical directions perpendicular to the X-ray beam propagation direction [31].

The diffracted X-rays propagated in an evacuated flight tube and were detected by an in-vacuum Eiger detector [43] placed at a distance of 5.260 m from the sample.

*Time resolved setup*

The time resolved measurements were performed with the pump-probe method, employing the X-ray flashes generated by the synchrotron light source as a probing signal. The time structure of the Swiss Light Source is designed to provide 70 ps wide (FWHM) X-ray pulses, defining the temporal resolution of our measurements, every about 2 ns (repetition frequency of 499.652 MHz). The magnetization dynamics in the sample were excited by an RF signal that was synchronized to the RF electron bunch accelerators of the synchrotron light source through an IQ modulator, which was then employed to shift the relative delay between the X-ray pulses and the exciting RF signal.

This method locks the excitation frequency and its phase to the synchrotron repetition frequency, or its higher order harmonics. If a single-bunch filling pattern were to be employed, classical pump-probe measurements could be carried out, which would allow the investigation of different frequency ranges, or pulsed excitations.

**Laminography measurements**

To measure 2D projections of the magnetization, resonant dichroic ptychography was used. The energy of the X-rays was tuned to the absorption edge of the magnetic element in question, in this case the $L_3$ edge of gadolinium with an energy of 7.247 keV, and ptychographic projections were measured with right and left circularly polarised light.

Ptychographic projections with a field of view of $7 \times 12.5$ μm$^2$ consisted of 818 diffraction patterns with a step size of 1 μm and an exposure time of 0.2s. The two dimensional projections of the magnetization shown in Figure 1b,c consist of an average of 25 single XMCD projections, which allowed for a sufficiently high signal-to-noise ratio 2D image of the magnetic structure.

For magnetic laminography, ptychographic projections were measured with both right and left circularly polarised X-rays at 144 angles equally spaced over 360°. Within each dataset, six of the projections were blanked by the coaxial cable carrying the RF excitation to the sample, meaning that the laminography reconstructions were performed using 138 projections of each polarization.

The time delays of the excitation field were measured non-sequentially to be sure that any observed dynamic behaviour would be independent of slight sample changes occurring over time due to heating or aging effects. The phases were measured in the following order:

$$\pi, 0, \frac{7\pi}{4}, \frac{3\pi}{4}, \frac{\pi}{4}, \frac{5\pi}{4}, \frac{\pi}{2}, \frac{3\pi}{2}$$

At times of $1 ns, 0 ns, 1.75 ns, 0.75 ns, 0.25 ns, 1.25 ns, 0.5 ns, 1.5 ns$, respectively.

Unfortunately, due to experimental errors it was not possible to obtain a complete reconstruction from the $\frac{\pi}{2}$ (0.5 ns) laminography dataset, which is why seven time-steps are presented in the manuscript rather than eight.

Each laminography measurement consisting of 144 projections took 7 hours including overhead, meaning that each magnetic laminography measurement required 14 hours in order to measure with both left and right circular polarization. Therefore, in total, seven time steps required ~4 days and 2 hours of continuous measurement and a total of 2016 projections.

**Laminography reconstruction (non-magnetic)**

The measured diffraction patterns were used to reconstruct the complex-valued projections $P_n(x,y)$. The reconstruction was performed using 300 iterations of the difference map method [44] and 300 iterations of the maximum likelihood optimization [45]. Reconstructed and unwrapped phase projections were aligned together in order to maximize their mutual consistency [46]. This is possible only because the magnetic properties of the sample do not affect the phase-based alignment procedure at the selected photon energy. Aligned and unwrapped phase projections were reconstructed by the filtered backprojection method. The laminography imaging geometry unavoidably results in missing information in the Fourier space, which is the so-called missing cone problem. Therefore, an iterative method enforcing physical constraints in the real space and measured projection in the Fourier space was used to fill the missing cone [31].

The non-magnetic reconstruction was used to estimate the location of the magnetic material, which is used as a constraint in the magnetic reconstruction detailed in the following section.

**Magnetic Laminography**

Magnetic laminography, as well as magnetic tomography, is a special case of arbitrary projection 3D magnetic imaging. For completeness we repeat here the general formulation for arbitrary projection magnetic tomography reconstructions [32] and later show the particular application to laminography.

The amplitude of the transmissivity, $A_n$, measured experimentally is related to the defined XMCD projection $P_n$ as follows:

$$A_n(x,y) = exp\left(\frac{-2\pi}{\lambda} P_n(x,y)\right)$$

After normalization and computation of the absolute value of the transmissivity, each projection is mathematically defined in terms of rotation matrices as:

$$P_n(x,y) = \frac{-r_e}{2\pi}\lambda^2 \left(\sum_k n_{at}^k \int \Im\{f_c^k(R^{(n)}r)\}dz \pm n_{at}^{mag} \int \Im\{f_m^{(1)}\}[R^{(n)\dagger}m(R^{(n)}r)]\cdot \hat{z}dz\right)$$

Where $r_e$ is the classical electron radius, $\lambda$ is the X-ray wavelength, $n_{at}^k$ and $n_{at}^{mag}$ are the atomic density of the $k$th element, and of the resonant magnetic element, respectively, and $\Im$ selects the imaginary part. Here, $m(R^{(n)}r)$ is the magnetization vector in the object coordinate system, and $f_c$ and $f_m^{(1)}$ are the electron density and XMCD scattering factors, respectively. The magnetic contribution originates only from the resonant element, given in the second term, whereas the electronic scattering is calculated by summing over all elements present, including the magnetic element.

During the magnetic laminography reconstruction, the current reconstructed components of the magnetization are used to compute projections and compare them to the data. In this calculation, the integrals are approximated by sums, and the projection is defined by:

$$\hat{P}_n(x,y) = \sum_k \{c[R^{(n)\dagger} m(R^{(n)} r)] \cdot \hat{z} + o(R^{(n)} r)\}$$

Where $m$ and $o$ are the estimated magnetic and non-magnetic structures, respectively, and $c$ is a constant that relates the XMCD signal to the magnetization, including the pixel size of the dataset. At each iteration of the reconstruction, the correctness of the reconstruction is estimated by calculating an error metric that compares the estimated dataset $\hat{P}_n(x,y)$ with the measured dataset $P_n(x,y)$. This error metric is defined as:

$$\varepsilon = \sum_n \sum_{x,y} [\hat{P}_n(x,y) - P_n(x,y)]^2$$

The reconstruction is then updated such that the error metric is reduced, and the reconstruction approaches the correct solution, by calculating the gradient of the error metric with respect to the different reconstructed variables. In this case, the gradients for the magnetization $m$ and the non-magnetic structure $o$ are analytically defined as:

$$\frac{\partial \varepsilon}{\partial m(R^{(n)} r)} = 2 \sum_n [\hat{P}_n(x,y) - P_n(x,y)] R^{(n)} \begin{bmatrix} 0 \\ 0 \\ 1 \end{bmatrix}$$

$$\frac{\partial \varepsilon}{\partial o(R^{(n)} r)} = 2 \sum_n [\hat{P}_n(x,y) - P_n(x,y)]$$

For magnetic laminography, the rotation matrices that define a rotation of the sample by $\varphi$ around the rotation axis, which is in turn at an angle $\theta_L$ with respect to the direction of propagation of the X-rays ($\hat{z}$), are given by:

$$R^{(n)\dagger} = \begin{pmatrix} 1 & 0 & 0 \\ 0 & \sin\theta_L & \cos\theta_L \\ 0 & -\cos\theta_L & \sin\theta_L \end{pmatrix} \begin{pmatrix} \cos\varphi_n & 0 & \sin\varphi_n \\ 0 & 1 & 0 \\ -\sin\varphi_n & 0 & \cos\varphi_n \end{pmatrix}$$

$$\begin{pmatrix} \sin\varphi_n & 0 & \cos\varphi_n \\ -\cos\theta_L \sin\varphi_n & \sin\theta_L & \cos\theta_L \cos\varphi_n \\ -\sin\theta_L \sin\varphi_n & -\cos\theta_L & \sin\theta_L \cos\varphi_n \end{pmatrix}$$

As a result, the gradients for the components of the magnetization are defined as:

$$\frac{\partial \varepsilon}{\partial m(R^{(n)} r)} = 2 \sum_n [\hat{P}_n(x,y) - P_n(x,y)] \begin{bmatrix} -\sin\theta_L \sin\varphi_n \\ -\cos\theta_L \\ \sin\theta_L \cos\varphi_n \end{bmatrix}$$

For $0 < \theta_L < 90°$, all three components of the gradient are non-zero, meaning that the measured projections are sensitive to, and contain information about, all three components of the magnetization. This in turn suggests that they can in principle be recovered with a dataset measured around a single rotation axis. To determine whether this is the case, we performed numerical simulations of magnetic laminography for a complex 3D magnetic structure containing vortices, domain walls and Bloch points, as in [21]. For the simulations, 360 XMCD projections were measured around the laminography rotation axis, and, in order to reconstruct the magnetic signal, 50 iterations of the arbitrary projection magnetic reconstruction algorithm (APMRA) were used. The laminography axis was varied between 0° and 90° to determine the optimal geometry for determining the 3D magnetization vector field with magnetic laminography.

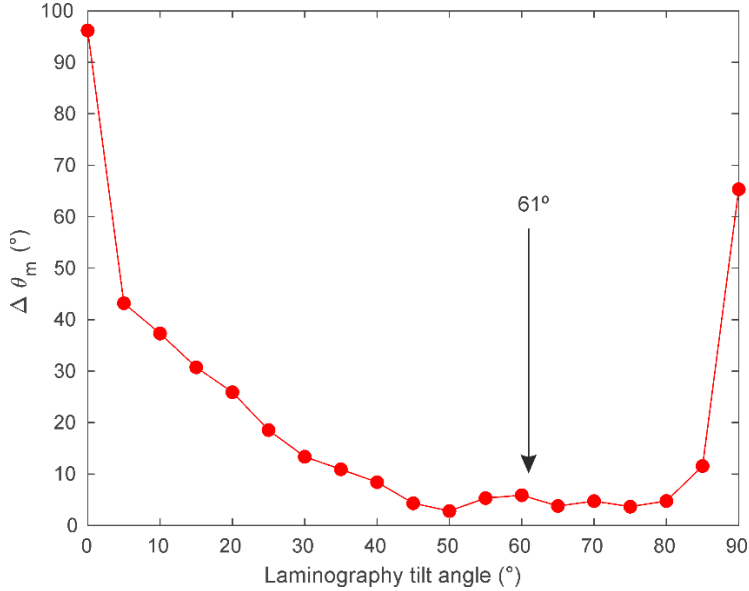

Figure S1: Numerical simulations of magnetic laminography for different laminography tilt angles show that there exists a range of angles (30°-80°) for which the 3D structure can be reconstructed with a high degree of accuracy. The accuracy of the reconstruction is determined by calculating the RMS error of the direction of the reconstructed magnetization with respect to the original structure ($\Delta\theta_m$). The tilt angle used in the experiment presented in this manuscript is 61°, which lies comfortably within this range and is indicated with an arrow.

The error in the reconstructed magnetic structure with respect to the original magnetic structure was defined as the error in the local magnetization 3D orientation and was calculated for a range of laminography tilt angles. This error in the direction of the reconstructed magnetization direction is shown in Figure S1, where it can be seen that a reasonably wide range of angles exists for which the 3D magnetic structure is reconstructed with a high degree of accuracy. In particular, for the angle used in this work $\theta_L = 61.2°$, 99% of all voxels have an error less than 6° in the direction of the magnetization in three dimensions.

*Magnetic laminography reconstruction*
The three-dimensional magnetic structure was reconstructed using 10 iterations of the GPU-implemented version of the gradient-based arbitrary projection reconstruction code described in [32] and made available at [47], using the rotation matrices described above. During the magnetic reconstruction, the magnetization was constrained to the magnetic material using a mask calculated from the non-magnetic laminography reconstruction. No additional regularisation of the magnetization was implemented.
Once reconstructed, the three components of the magnetization were filtered using a Hanning filter with a cutoff spatial frequency of 1/6 with respect to the maximum spatial frequency in order to remove high frequency noise, and the magnetization was then normalised by its magnitude to obtain a magnetic vector field of uniform amplitude.

The magnetic reconstruction for each time-step was performed independently using the same reconstruction parameters.

**Spatial resolution and accuracy of magnetic reconstruction:**
The full-period spatial resolution of the magnetic reconstruction was estimated by taking a line profile of the $m_x$ and $m_z$ components across the vortex domain wall and measuring the 25%-75% edge sharpness, as shown in Figure S2, which was found to be 50 nm. Micromagnetic simulations show that at the surface of a sample, the vortex core in such a system is of the order of 20-30 nm, which increases rapidly as one moves into the bulk of the disc. Considering that the measured edge sharpness is a convolution of the size of the vortex core and the spatial resolution, we therefore estimate the half-period spatial resolution of the magnetic reconstruction to be around 50 nm.

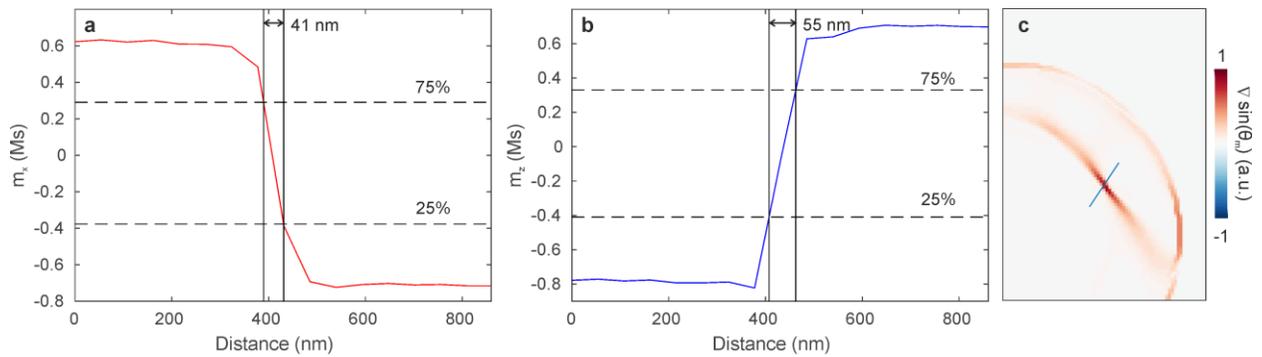

Figure S2: Line profiles of (a) the $m_x$ and (b) the $m_z$ components of the magnetization across the vortex domain wall. The 25%-75% edge sharpness is measured to be on average 50 nm, indicating a half-period spatial resolution of 50 nm. c) The line along which the profile is measured is drawn in blue

A further estimate of the half-period spatial resolution was also estimated by calculating the Fourier Shell Correlation (FSC) of the three components of the magnetization and the in-plane vector fields in the three Cartesian planes, shown in Figure S3. The FSC indicated an average spatial resolution of 184 nm, 360 nm and 180 nm for the $m_x$, $m_y$ and $m_z$ components respectively. The difference between the edge sharpness and FSC estimated resolution can be explained by the spatial sparsity of the data, i.e. the small number of features with high spatial frequencies present in the sample., and the resulting lack of high frequencies present, which leads to underestimation of the spatial resolution based on FSC [48].

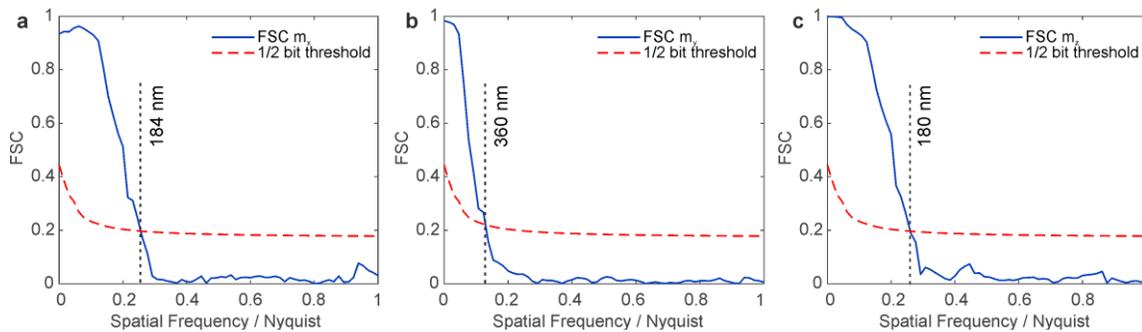

Figure S3. Fourier Shell Correlation of (a) the $m_x$, (b) the $m_y$ and (c) the $m_z$ components of the magnetization. An estimate of the spatial resolution is obtained using the ½ bit threshold [36] (red dashed line) and is found to be 184 nm, 360 nm and 180 nm for the $m_x$, $m_y$ and $m_z$ components of the magnetization, respectively

In the time-resolved measurement, we are able to track displacements of the vortex domain wall with an accuracy significantly higher than the spatial resolution. In this case, we track oscillations in the domain wall location with an amplitude of 45 nm and 74 nm for the vortex domain walls with negative and positive vorticity, respectively. This sub-spatial resolution tracking of the domain wall is possible as it is dependent on the signal-to-noise ratio of the reconstruction, and not the global or local spatial resolution.

**Three-dimensional magnetic structure**
For reference, the heights for which the magnetization is shown in Figure 2a-d, and the 500 MHz modes in Figure 3a-d, are indicated on the three-dimensional rendering of the 500 MHz modes in Figure S4.

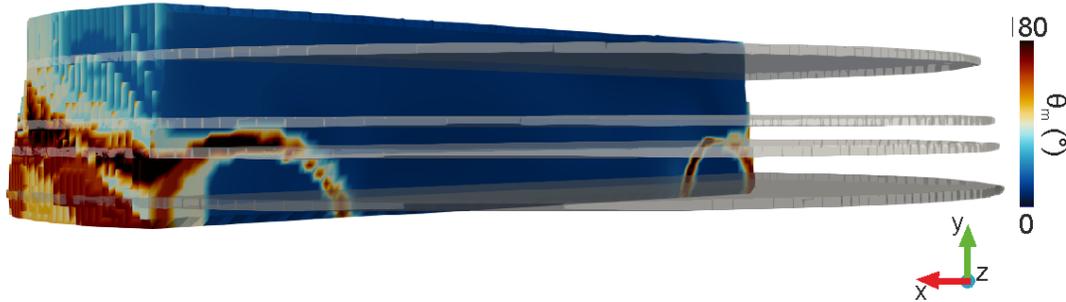

Figure S4. The heights at which the magnetization and dynamics are shown in Figs 2a-d and 3a-d are indicated by semi-transparent discs, superimposed on the three-dimensional rendering of the magnetization modes shown in Figure 3e.

**Data Analysis: Calculation of vortex domain wall displacement**
The displacement of the vortex domain wall was calculated using two different procedures that both make use of the sub-pixel image registration algorithm described in [49]:

*Vortex-tracking based on $\nabla \theta_m$:*
The divergence of the direction of the in-plane magnetization was calculated for a region of the sample surrounding the vortex core of the domain wall, as in Figure 4a for the case of the vortex domain wall with positive vorticity. As an example, the divergence of $\theta_m$ is shown in Figure S5, where one can see that at the core of the vortex, there is a sharp peak in the gradient of $\theta_m$, with a FWHM of only a few pixels. The location of the vortex core for each time-step was determined by aligning the image of $[\nabla sin(\theta_m)]_t$ with a reference two-dimensional Gaussian with a FWHM of 1 pixel (46.8 nm). The same analysis was performed for all slices of the three-dimensional magnetic structure in which the vortex domain wall was present (the lower ~500 nm of the disc), and the motion of the vortex domain wall as a function of height was determined.

*Vortex-tracking based on $sin(\theta_m)$:*
While using $\nabla \theta_m$ worked well for tracking the vortex with positive vorticity shown in Figure 4, when applied to the second vortex domain wall with negative vorticity, the tracking was found to be more susceptible to noise due to the more complex structure of the domain wall. To reduce the noise, we chose to align images with signal that was not limited to the vortex core, but was present throughout the image. For this, we calculated $sin(\theta_m)$, shown in Figure S5b, which had a non-zero signal throughout the image, and does not exhibit phase-wrapping, i.e. the sharp transition from $0 \rightarrow 2\pi$, as it does for $\theta_m$.

The components of the motion of the vortex domain wall parallel and perpendicular to the long axis of the domain wall were calculated and compared. In general, for both domain walls, the motion perpendicular to the long axis of the domain wall could be identified as oscillatory. However, the motion parallel to the domain wall was much noisier, and no clear oscillatory motion could be determined. The challenges in determining the motion parallel to the long axis of the domain wall are likely due to the extreme ellipticity of the vortex core due in turn to the anisotropy in the system. In this case, it is likely that the much smaller displacement of the core parallel to the long axis of the domain wall is below the current accuracy of this measurement.

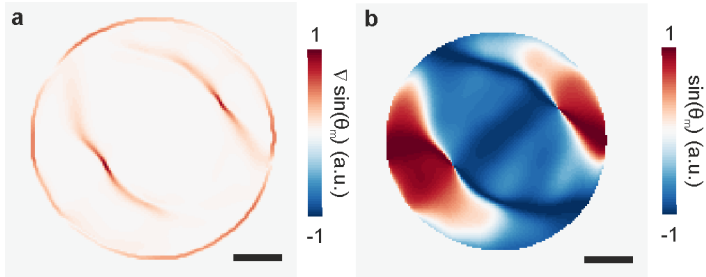

Figure S5. Images of a) $\nabla sin(\theta_m)$ and b) $sin(\theta_m)$ of the lowest slice of the magnetization. One can see that for $\nabla sin(\theta_m)$ in a), the signal is limited to the domain walls, whereas for $sin(\theta_m)$ the signal is non-zero throughout the structure.

The motion of the second vortex domain wall perpendicular to its long axis, indicated by grey arrows in Figure S6, is given as a function of height in Figure S6 and here one can identify a similar periodic temporal behaviour to the first vortex domain wall shown in Figure 4, albeit with a smaller amplitude of (45±13) nm, and a phase shift of (0.75±0.25) ns with respect to the first vortex domain wall.

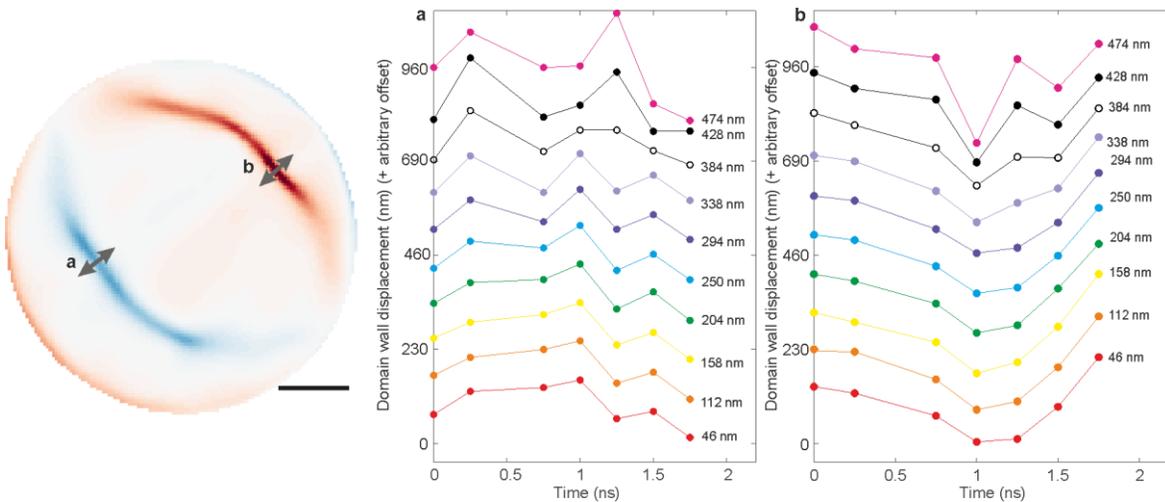

Figure S6. Displacement of the vortex domain wall with negative vorticity (a, blue in Figure 4) and positive vorticity (b, red in Figure 4) in the direction perpendicular to the long axis of the domain wall (indicated by grey arrows) as a function of time for different heights within the structure. The oscillatory behaviour is clearest for both vortices at a height of 46 nm (the bottom of the structure) and is noisier at the top. When comparing the two vortices, the oscillations appear to be consistently out of phase through the height of the sample, as shown for $h = 46\ nm$ in Figure 4c. Scale bar is 1 μm.